\documentstyle[prb,aps,twocolumn,floats]{revtex}  
  
\begin{document}  
  
\title {\bf On the thermal broadening of a quantum critical phase
transition
}  
  
\author{ P.T. Coleridge and P. Zawadzki}  
  
\address  
{Institute for Microstructural Sciences, National Research  
Council, Ottawa, Ontario, K1A OR6, Canada\\ }  
  
\date{16 March 1999}  
  
\maketitle  
\begin{abstract}  
{
 The temperature dependence of an integer Quantum Hall effect transition is
studied in a sample where the disorder is  dominated by short-ranged 
potential scattering. At low temperatures the results are consistent with a 
$(T/T_0)^{\kappa}$ scaling behaviour and at higher temperatures by a 
linear dependence similar to that reported previously in other material 
systems. It is shown that the linear behaviour results from thermal broadening
 produced by the Fermi-Dirac distribution function and that the 
temperature dependence over the whole range depends only on the scaling 
parameter T$_0^{\kappa}$.
 
}  
\end{abstract}  

     In a recent paper \cite{ptc1} (referred to hereafter as I) it was
demonstrated, for a sample in which the disorder is dominated by short-
ranged potential scattering, that the 2-1 integer quantum Hall transition
was well described by a scattering parameter 
  
\begin{equation}  
             s     =  \exp [ - \Delta \nu / \nu_0 (T)] ,    
\label{eq1}  
\end{equation}

\noindent where $\Delta \nu$ (= $\nu - \nu_c$) is the deviation of the
filling factor from the critical value ($\nu_c$) and $\nu_0 (T)$, at least
in the low temperature limit, obeys a scaling law $(T/T_0)^{\kappa}$ with
$\kappa \approx  3/7$. It was argued that this description, which is
characteristic of a quantum critical phase transition \cite{DasSarma}, is
also appropriate for transitions into the Hall insulator state and for the
B = 0 metal-insulator transition.

     By contrast, Shahar et al \cite{Shahar1} (hereafter referred to as II)
studying the Hall insulator transition in a variety of GaAs and InGaAs
based samples, find the same exponential dependence on filling factor but a
logarithmic slope $\nu_0 (T)$ that appears to vary as $ \alpha T + \beta $
rather than exhibiting $T^{\kappa}$ scaling behaviour. The ratio $\beta /
\alpha $, defines a temperature that is found to be characteristic of the
material system.

\begin{figure} [t]  
\vspace*{7.3cm}  
\includegraphics{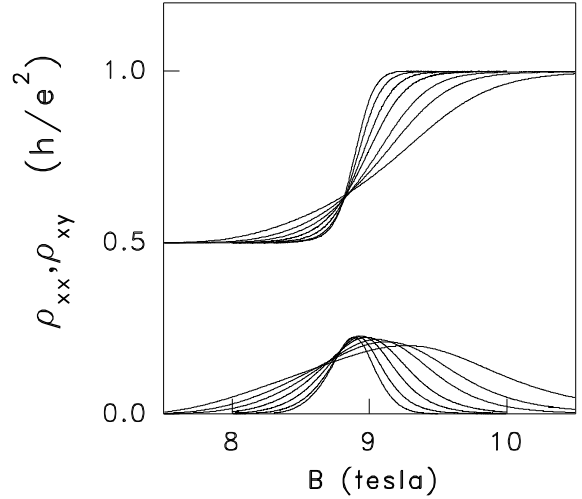}  
\caption  
{Resistivity data for the $\nu$ = 2 to 1 IQH effect transition 
at temperatures of 60, 100, 150, 210, 300, 400 and 600mK. }  
\label{fig1}  
\end{figure} 
  
\begin{figure} [t]  
\vspace{6.8cm}  
\includegraphics{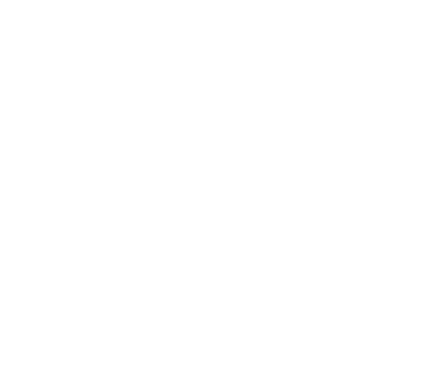}  
\caption  
{ Conductivity data, plotted as a function of filling factor $\nu$,
obtained by inverting the resistivities shown in Fig.1 .
}  
\label{fig2}  

\vspace{5.5cm}  
\includegraphics{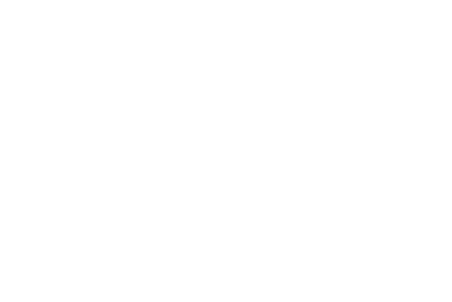}  
\caption  
{Scattering parameter s, defined in the text, derived from the
$\sigma_{xy}$
data shown in Fig. 2, plotted on a logarithmic scale.
}  
\label{fig3}  
\end{figure}  

     New results are presented here for the same p-SiGe sample studied in I
but measured in more detail and over a wider temperature range. The same
approximately linear dependence of  $\nu_0 (T)$ seen in II is observed. It
is attributed, not to a new transport regime, but rather to the situation
where the behaviour is dominated by thermal broadening associated with the
Fermi-Dirac distribution function.

     Figure 1 shows resistivity components, $\rho_{xx}$ and $\rho_{xy }$
around the 2-1 quantum Hall transition, measured for a range of
temperatures between 60 and 600mK. The sample and experimental procedures
have been described previously \cite{ptc1}. Although a fixed point can be
seen in the Hall data the corresponding point in the 
the longitudinal resistivity is
progressively smeared out as the temperature is raised. This behaviour can
be understood more clearly when the data is inverted to obtain conductivity
components (Figure 2). The Hall conductivity shows a well defined fixed
point, at $\sigma_{xy}$ = 1.5 (in units of  e$^2$/h) but the peak value of
$\sigma_{xx}$, which should correspond to a temperature independent
critical point, deviates progressively from the expected value of 1/2 as
the temperature increases. It should be noted, however, that there is a
high degree of reflection symmetry, ie $\sigma_{xx}(\Delta \nu)=
\sigma_{xx}(-\Delta \nu)$.

\begin{figure} [t]  
\vspace{11.8cm}  
\includegraphics{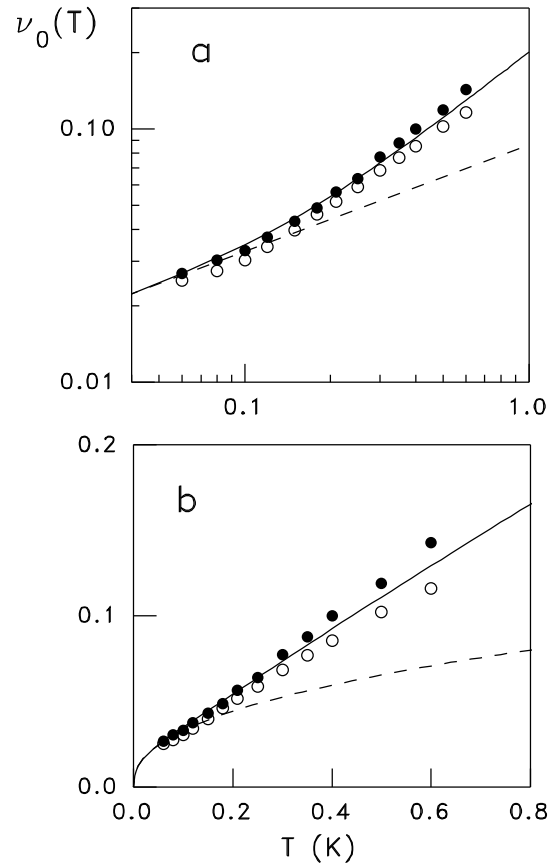}  
\caption  
{Logarithmic slope, $\nu_0 (T)$, derived from $\sigma_{xy}$ values (solid
points) and $\sigma_{xx}$ values (open points) plotted against temperature.
using (a) log-log scales and (b) linear scales.  In both cases the solid
line represents the results of the thermal broadening calculation presented
in the text and the dashed line a $T^{.43}$ dependence.
}  
\label{fig4}  
\end{figure}

     Following the approach outlined in I the scattering parameter s can be
extracted from this data according to
  
\begin{equation}  
        \sigma _{xx} = 2 \sigma^{pk}(T) s / (1 + s^{2}), \; \; \; \; \;     
             \sigma _{xy} = 2 - s^{2} / (1 + s^{2}).
\label{eq2}  
\end{equation}

\noindent A prefactor $2 \sigma^{pk}$, with a weak temperature dependence, 
has been introduced for $\sigma_{xx}$ to account for the deviations from 
exact critical behaviour. As discussed in I these are attributed to the 
finite range of the impurity scattering potential. Although this is small,
corresponding to predominantly large angle scattering, it is nevertheless
finite so the momentum weighting factor, (1-cos$\theta$), plays a role in
the transport processes.

\begin{figure} [t]  
\vspace{6.0cm}  
\includegraphics{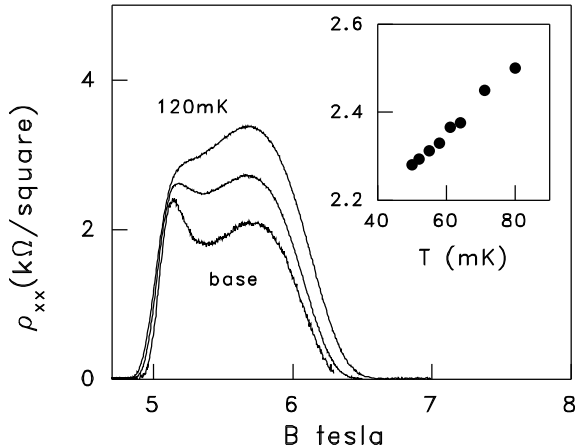}  
\caption  
{Feature in the $\rho_{xx}$ data, between filling factors 3 and 2,
associated with a spin transition. Nominal ( i.e. mixing-chamber)
temperatures are 120mK, 60mK and ``base'' (estimated to be about 30mK).
Inset: peak height of
the feature as a function of mixing chamber temperature.
}  
\label{fig5}  
\end{figure}  

     Values of s deduced from  $\sigma_{xy}$ are shown in figure 3: very
similar results are obtained from $\sigma_{xx}$.  In agreement with earlier
work \cite{ptc1,Shahar1,Pan2,xx3} the field dependence near the critical
point is well described by eqn. 1 and values of $\nu_0(T)$,  derived from
the slopes at $\Delta \nu = 0$, are shown in fig. 4. Also shown are values
obtained from the $\sigma_{xx}$ data, deduced from the width of the peak
measured at half-maximum height. In agreement with the results reported in
II the temperature dependence of $\nu_0 (T)$, over this relatively
large range, is not given by a $T^{\kappa}$ scaling behaviour but is better
described by a linear variation of the form $\alpha T + \beta$. For the
data shown the characteristic temperature given by $\beta/\alpha$  is .08K,
very similar to the values found in II for GaAs/GaAlAs samples. At the
lowest temperatures, however, the data is consistent with an asymptotic
approach to the expected scaling behaviour with $\kappa \approx 3/7$.

     To support the identification of the low temperature behaviour with
scaling it is important to establish experimentally that change in slope
around T = 0.1K seen in fig 4a is a genuine feature and not just the result
of temperature saturation in the sample produced by spurious heating
effects. Fortunately, there is available here an independent measure of the
temperature of the holes. Figure 5 shows $\rho_{xx}$ data between filling
factors $\nu$ = 3 and 2 at dilution refrigerator temperatures of 120mK, 60mK 
and ``base'' temperature (about 30mK). In addition to the 3-2 integer quantum
 hall transition at 5 tesla there is extra structure, at higher fields,
 that is strongly dependent on both temperature and measuring current. It 
has been established by activation measurements that this corresponds to a
paramagnetic/ferromagnetic phase transition induced by exchange enhancement
of the spin splitting \cite{ptcssc}.  As shown in the inset the amplitude
of this peak varies approximately linearly with
temperature down to 50mK, the lowest temperature for which the Ge
resistance thermometer, mounted in the mixing chamber, is calibrated. As
the main graph shows, the peak height continues to decrease as the mixing
chamber temperature is further lowered. This provides some confidence that
at the lowest temperature shown in fig 4 (60mK) the deviation of
temperature of the holes in the sample from that measured by the Ge
thermometer is at most a few mK.

     It is argued therefore that at the lowest temperatures 
scaling as $T^{\kappa}$ is observed but as the temperature 
is raised this is overwhelmed by the
approximately linear dependence. The linear increase is attributed to
thermal broadening associated with the Fermi-Dirac distribution function
f(E). This is qualitatively reasonable: in the absence of localisation the
Landau level width is of order 5K (given for example by a Gaussian line
shape \cite{Raikh&S} with a width $ (e \hbar / m^{\ast}) (B / 2\pi
\mu_q)^{1/2}$ where $\mu_q$, the quantum mobility, is 1.5 m$^2$/Vs ). When
localisation becomes important the small band of itinerant states near the
centre of the Landau level that dominates the transport behaviour 
will have a characteristic energy width roughly
an order of magnitude smaller than this,  comparable with the width of 
the Fermi function ($- \partial f/\partial E$) at temperatures of order
300mK. This ``classical'' effect was in fact noted in some of the earliest
results presented on localization and scaling \cite{HPWei}.  

     The thermal broadening can be calculated using the standard
approach \cite{Landau}. For a conductivity with an energy dependence
given by $\sigma (E)$, the temperature variation associated with thermal
broadening is given by 
  
\begin{equation}  
             \sigma (T)   =  \int \sigma (E) (- \partial f/\partial E) dE.  
\label{eq3}  
\end{equation}  

\noindent If the scattering parameter s is determined by quantum critical
behaviour, ie if $\nu_0 (T) = (T/T_0)^{\kappa}$ in eqn. 1, the energy
dependence of s for $E \approx E_c$ is 
  
\begin{equation}  
      s(E)     =  \exp [ - (E -E_c) (T_0/T)^{\kappa}/ \gamma],
\label{eq4}  
\end{equation}

\noindent where $E_c$ is the energy associated with the critical point and
$\gamma$, proportional to Landau level width, is used to relate $\Delta
\nu$ to $E_F -E_c$. 

     Using eqns 2-4, with the $\sigma_{xy}$ conductivity, gives the solid
line shown in fig4. This result is insensitive to the value of $(E_F -
E_c)/\gamma$  so there is only one adjustable parameter in the calculation,
$T_{0}^{\kappa}$, which is chosen to give agreement in the low temperature
asymptotic regime.  The calculation then automatically gives the correct
behaviour for higher values of T. In light of the relatively crude model
used to define the energy dependence of s(E) the agreement with experiment
is very good. Similar behaviour is also observed when $\sigma_{xx}$ is used
as a basis for the calculation although there is some difficulty then in
knowing how, precisely, to deal with the deviation of the peak value (cf
eqn. 2) from 0.5e$^2$/h. The thermal broadening not only gives the linear
increase of $\nu_0 (T)$ but also a temperature dependent reduction of the
peak height, of the correct order of magnitude. A proper treatment of this
effect, however, requires the inclusion of the momentum 
weighting term (1-cos$\theta$) .

     The good agreement between the calculated and experiment values of
$\nu_0 (T)$ therefore confirms that the linear dependence should be
associated with thermal broadening and does not represent a new transport
regime. The characteristic temperature $\beta/\alpha$ is not a new
temperature scale but rather just depends on T$_0$. Indeed, it provides a
means of determining this parameter even when the low temperature scaling
regime happens to be experimentally inaccessible.

     For the data presented here T$_0$ is 290K and the ratio of the two
temperatures is about $2.7\times 10^{-4}$. Similar values also 
pertain to the GaAs/GaAlAs data shown in II. For the InGaAs/InP 
data shown in II, where $\beta / \alpha$ is about 0.5K, scaling behaviour
is not seen. However, for a similar sample \cite{HPWei2} T$_0$ is 
approximately 3400K with a corresponding  ratio of about 
$1.5\times 10^{-4}$,  of the same order of magnitude. This suggests
therefore, that thermal broadening provides a means whereby these two 
apparently conflicting experimental results, obtained in nominally
identical samples, can be reconciled.

     In summary, for a sample where the disorder is dominated by 
short-ranged potential
scattering, so the transport data is an accurate reflection of the quantum
critical processes, the temperature dependence of the 2-1 integer
 quantum Hall transition is characterised by two regimes. In a low 
temperature regime the behaviour is consistent with scaling
having the expected exponent ( $\kappa = 3/7$), while 
at higher temperatures, in agreement with earlier work, 
there is a linear dependence. It is suggested, and confirmed by calculation, 
that the linear dependence should be attributed to thermal broadening from
the Fermi-Dirac distribution function. 

     Acknowledgements: We would like to thank R. Williams and Y. Feng for
the growth and fabrication of the sample.


\end{document}